\documentclass{pazh_engl}
\usepackage{epsf}
\usepackage{graphicx}
\usepackage{lscape}
\usepackage{graphicx,natbib}
\bibpunct{(}{)}{;}{a}{}{,}

\begin{document}

\title{New nearby AGNs from all sky surveys of INTEGRAL and RXTE
  observatories}

\author{I.F.Bikmaev \inst{1,2}, R.A. Sunyaev\inst{3,4}, M. G.
  Revnivtsev\inst{3,4}, R. A. Burenin \inst{3}}

\institute{
  Kazan State University, ul. Kremlevskaya 18, Kazan, Russia
  \and
  Academy of Sciences of Tatarstan, ul. Baumana, 20, Kazan, Russia
  \and
  Space Research Institute (IKI), ul. Profsoyuznaya 84/32, Moscow, Russia
  \and
  Max-Planck-Institut f\"ur Astrophysik, Garching, Germany
}

\bigskip
\bigskip

\abstract{We present first results of a campaign of optical identifications
  of X-ray sources discovered by RXTE and INTEGRAL observatories during
  their all sky surveys. In this work we study six newly discovered nearby
  active galactic nuclei at $z<0.1$. The optical spectrophotometric data
  were obtained with Russian-Turkish 1.5-m telescope (RTT150). We present
  their redshifts and main parameters of brightest emission lines.}

\date{5 november 2005}
\titlerunning{NEW NEARBY AGNs FROM INTEGRAL AND RXTE SURVEYS}
\authorrunning{BIKMAEV ET AL.}

\maketitle

\section*{INTRODUCTION}

During last decade a significant progress in understanding the properties of
active galactic nuclei (AGNs) and their evolution was achieved. In
particular, it was shown that the rise of volume density of low luminosity
AGNs occurred at much lower redshifts than that of high luminosity AGNs
\citep[see e.g.,][]{steffen03,ueda03}.

Observations show that a large fraction of cosmic X-ray background is
contributed by AGNs with high intrinsic photoabsorption \citep[see
e.g.,][]{hasinger04}, while fraction of absorbed AGNs rises with the
decrease of their X-ray luminosity \citep{ueda03,sr04}.

In frames of widely accepted ``unification scheme'' of AGNs \citep[see
e.g.,][]{antonucci85,krolik88,setti89,antonucci93} the absorbtion observed
in X-ray and optical spectra of AGNs is explained as a result of obscuration
by molecular torus around the central black hole of AGN. In order to probe
different theories of X-ray and optical absorption (i.e., theories of AGN
unification) and in order to obtain accurate estimates of AGN volume density
in local Universe ($z<0.1$) one need to study large sample of low redshift
AGNs. Unfortunately, it is not possible to obtain this sample in deep
surveys of extragalactic fields, which were done recently with last
generation X-ray telescopes \citep[e.g.][]{brandt01,hasinger01}, due to
their small sky coverage.

It is very hard for X-ray telescopes to cover a large fraction of the sky
due to their small fields of view. Until recently practically the only
available X-ray survey at the energies above 2 keV was HEAO1/A2 all sky
survey \citep{piccinotti82}. Recently completed all sky survey of RXTE
observatory (XSS - RXTE Slew Survey, \citealt{mikej04}) allowed to improve
the sensitivity and considerably increased the X-ray sample of local AGNs
\citep{sr04}.  However, due to not very hard energy band of the RXTE survey
(3--20 keV) its sensitivity to highly absorbed AGNs ($\log(N_{\rm H}L)>23$)
is low, which introduce a considerable bias in the sample. Large solid angle
surveys in hard X-rays of INTEGRAL and SWIFT observatories (Revnivtsev et
al. 2006, in preparation; \citealt{gehrels04,markwardt05}) allow one to
avoid this bias.

At the present time observations of INTEGRAL satellite \citep{winkler03}
covered a significant part of the whole sky and allowed one to discover a
set of AGN candidates (see e.g. \citealt{sazonov05}; Revnivtsev et al. 2006,
in preparation). Optical identifications of these AGN candidates are
complicated in many cases because of insufficiently accurate positions of
these sources provided by INTEGRAL telescopes (typically 2--3$^\prime$). It
is especially difficult to identify sources which have no counterparts in
other catalogs which have more accurate positions, e.g.\ ROSAT all sky
survey catalog \citep{voges99}.

The number of nearby ($z<0.1$) X-ray bright ($L_{\rm x}>10^{42-43}$ erg/s)
AGNs is not very large. For example, the most sensitive up to date all sky
X-ray survey (XSS) contains only $\sim$75 AGNs with redshifts $z<0.1$.
Therefore, even few nearby AGNs added to this sample are of value.

Recently we (in collaboration with S.Sazonov, E.Churazov and A.  Vikhlinin)
initiated a campaign of optical identification of sources from RXTE and
INTEGRAL all sky surveys. In this Letter we present first results obtained
during this project. The optical data were obtained with Russian-Turkish
1.5-m Telescope (RTT150, T\"UBITAK National Observatory, Antalya, Turkey).
For the first series of observations we selected only sources with
relatively accurate positions, determined with ROSAT, EINSTEIN, SWIFT or
CHANDRA \citep{sazonov05} observatories. Six studied XSS and IGR sources
were identified with AGNs in nearby galaxies.

\section*{DATA ANALYSIS}

As a first set of X-ray sources we have selected RXTE and INTEGRAL objects
of the northern sky (Dec$_{\rm J2000}>-30^\circ$) which have the best
localization accuracy because of observations of EINSTEIN, ROSAT, CHANDRA or
SWIFT satellites. List of studied sources is presented in Table~\ref{list}.

Observations were performed with RTT150 during August 8--13, Sept.  5--10
and Oct. 4--9, 2005 using low resolution spectrometer TFOSC and
CCD-photometer.

\begin{table*}
  \caption{List of identified AGNs and their main parameters.\label{list}}
  \tabcolsep=1.mm
  \begin{tabular}{lcccccccl}
    Source Name&R.A.~~~~~~~~  Dec. &$R_c$&z&$\log L_{\rm O[III]}$&$\log
    L_x$&FWHM$_{\rm H_\alpha}$,&Type&Comment\\
    & (J2000) & & & & & km/s\\
    \hline
    XSS J05054$-$2348 &  05 05 45.7 $-$23 51 14 &16.6&  $0.0351(2)$&41.1&43.6&$<685$& Sy2 &2MASX J05054575$-$2351139\\
    XSS J16151$-$0943 &  16 15 19.1 $-$09 36 14 &14.8&  $0.0650(2)$&42.0&43.9&$1600$& Sy1 &1RXS J161519.2$-$093618\\
    IGR J18559$+$1535 &  18 56 00.6 +15 37 58 &16.6&  $0.0838(2)$&42.2&44.7&$3200$& Sy1 &2E 1853.7+1534\\
    IGR J19473$+$4452 &  19 47 19.4 +44 49 43 &17.2&  $0.0532(2)$&41.4&43.5&$<685$& Sy2 &2MASX J19471938+4449425\\
    IGR J21277$+$5656 &  21 27 44.4 +56 56 35 &16.6&  $0.0144(2)$&41.6&43.2&$1600$& Sy1 &1RXS J212746.7+565606\\
    XSS J21354$-$2720 &  21 34 45.1 $-$27 25 56 &15.8&  $0.0670(2)$&41.9&43.8&$1190$& Sy1.5?&1RXS J213445.2$-$272551\\
    \hline
  \end{tabular}

  \medskip
  X-ray fluxes were calculated for 2-10 keV energy band, assuming power law shapes of AGN spectra ($\Gamma=1.8$).
  Observed fluxes were taken from RXTE survey \citep[3--20~keV,][]{mikej04} or
  measured with CHANDRA \citep[2--10~keV,][]{sazonov05} or INTEGRAL
  (17--60~keV, Revnivtsev et al. 2006, in preparation).

\end{table*}

\begin{table*}
  \caption{Parameters of some lines in AGN spectra\label{lines}}
  \begin{tabular}{l|cc|cc|c|cc}
    &\multicolumn{2}{c}{$H_\gamma, \lambda 4340$}&\multicolumn{2}{c}{$H_\beta, \lambda 4861$}&{$O[III], \lambda 4959/\lambda 5007$}&\multicolumn{2}{c}{$H_\alpha, \lambda 6563$}\\
    Source      & FWHM, \AA &EW, \AA&FWHM, \AA &EW, \AA&EW, \AA&FWHM, \AA &EW, \AA\\
    \hline
    XSS J05054-2348 &$10$&$18$&$<15$&$18$ &$34/102$&$<15$&$80$\\
    XSS J16151-0943 &$26$&$30$&$33$ &$94$ &$16/52$ &$35$&$398$\\
    IGR J18559+1535 &$-$ &$-$ &$69$ &$115$&$16/47$ &$70$&$542$\\
    IGR J19473+4452 &$<10$&$4$&$<14$&$7$  &$24/69$ &$<15$&$32$\\
    IGR J21277+5656 &$20$&$52$&$37$ &$169$&$28/77$ &$35$&$515$\\
    XSS J21354-2720 &$12$&$18$&$17$ &$40$ &$35/107$&$26$&$160$\\
    \hline
  \end{tabular}

  \medskip
  Spectral resolution in these observations was $\Delta
  \lambda\sim15$\AA. Widths of hydrogen lines are given with corrections to
  instrumental widening. Accuracy of determination of lines equivalent widths
  ($EW$) $\sim$10-15\%.

\end{table*}

\subsection*{Spectrometer TFOSC}

The T\"UBITAK Faint Object Spectrograph and
Camera\footnote{http://astroa.physics.metu.edu.tr/tug/tfosc.html} (TFOSC) is
a direct imaging and low-to-medium resolution spectrometer, similar to DFOSC
and other instruments of this series, build at Copenhagen University
Observatory. It was build and delivered to RTT150 telescope only recently.
In this Letter we present first scientific results from spectrometric data
obtained with this instrument.

In order to minimize the losses of light we have selected the slit width 100
$\mu$m (corresponding to 2.6$''$ angular size on the sky). For this study we
used the grism \#15, which provides the maximal light efficiency and the
widest spectral range (3300--9000~\AA). Obtained spectral resolution is
12--15~\AA\ (600--800~km/s)

Grism \#7 (spectral range 3900--6750~\AA) provides better spectral
resolution ($\sim$7 \AA), but for objects with redshifts $z>0.03$ important
part of the spectrum with group of lines $H_\alpha$ and $[N II],
\lambda\lambda 6548,6584$~\AA\ would be lost which will require additional
observations with grism \#8 (5800--8000~\AA). Spectra with higher spectral
resolution (grisms \#7 and \#8) of some objects will be presented in
separate paper.

Since studied objects are not too faint (see Table~\ref{list}),
during our observations we used 5--10 min exposures, depending on a
source brightness. For every object we obtained 2--6 individual
spectra. Inside series of spectral images the calibration spectra of
neon and helium calibration lamps were taken. Every night we also
obtained spectra of spectrophotometric standards at corresponding
zenith angles.

Analysis of obtained spectral data was done with modified version of DECH
package \citep{dech}. Overall accuracy of absolute wavelengths calibration
is $\sim$1~\AA. For subsequent photometric calibration of obtained spectra
we used our observations of spectrophotometric standards. Additional
correction for interstellar reddening was done using the neutral hydrogen
map of \cite{dickey90}.

\begin{figure*}[t]
\includegraphics[width=\textwidth]{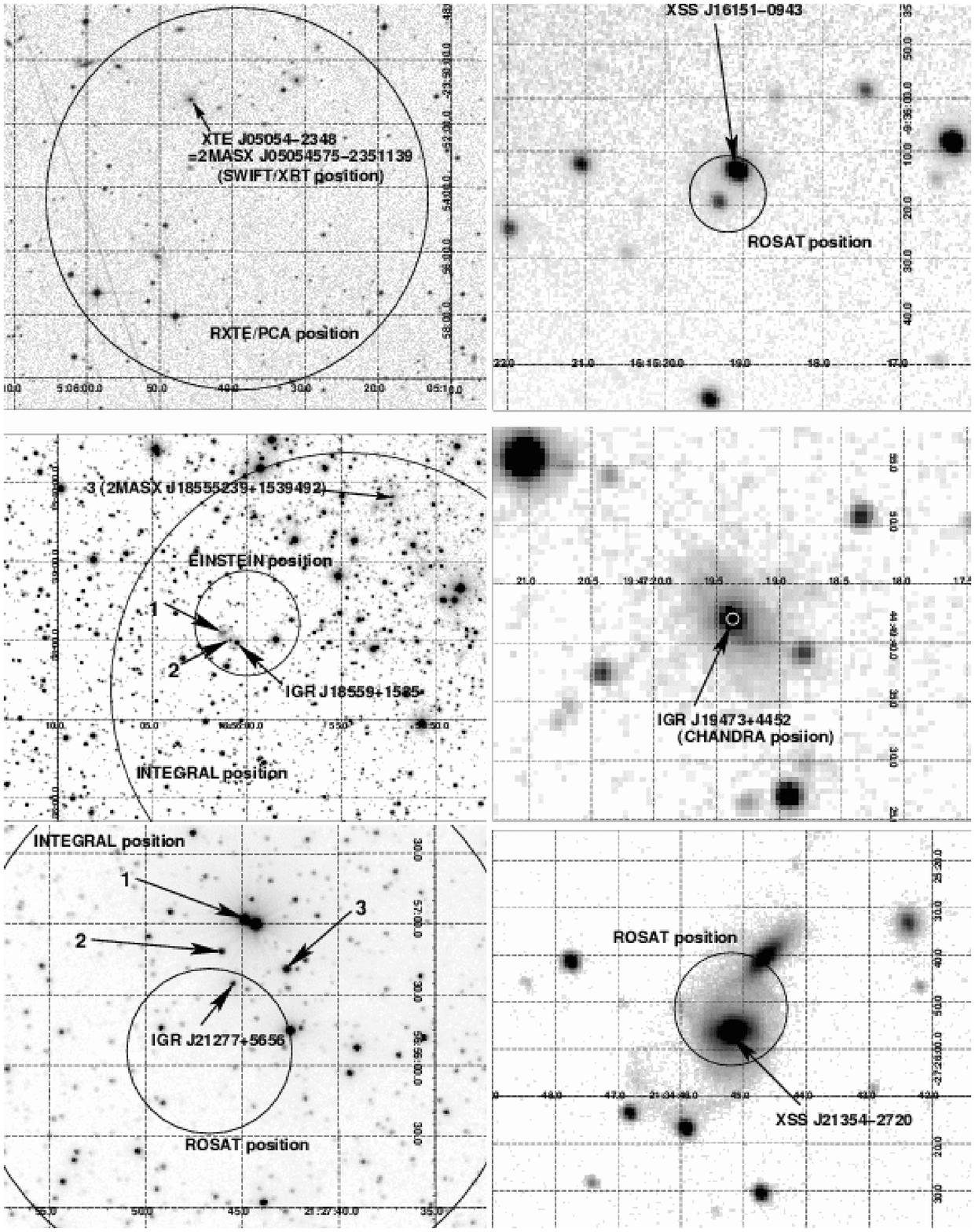}
 
\caption{RTT150 images of the sky in $Rc$ filter around
unidentified X-ray sources of RXTE and INTEGRAL all sky
survey\label{fcharts}} \label{fcharts}
\end{figure*}

\section*{RESULTS}

We have obtained deep (limiting magnitude $Rc~23.5$ mag) images of
localization regions of AGN candidates (see Fig.~\ref{fcharts}).
Circles on Fig.~\ref{fcharts} show different localizations of these
sources obtained by RXTE, ROSAT, CHANDRA, INTEGRAL and SWIFT
satellites. In some cases localization accuracy was high enough to
assure unambiguous identification of optical counterpart of the
X-ray source.  If more than one relatively bright ($Rc\la16$)
optical object was found within localization uncertainty radius we
obtained optical spectra of several optical objects. These objects
are marked with numbers in Fig.~\ref{fcharts}.

For all these X-ray sources we found optical objects with bright emission
lines of $H_\alpha$ and $O[III]$ with redshifts $z>0$ which clearly
indicates the presence of nearby galaxies.

Width of hydrogen lines and ratio of fluxes of lines $H_\alpha,
H_\beta, O[III]$ and $N[II]$ reveal the presence of an active
nucleus in the galaxies (shown by arrows in Fig.~\ref{fcharts}).
Some parameters of obtained spectra are presented in
Tables~\ref{list} and \ref{lines}. Spectra, normalized on continuum
flux are presented in Fig.~\ref{spectra}.

Study of deep images of AGNs showed that in all cases we detected extended
object around the AGN --- their host galaxies. Angular sizes of galaxies at
the sensitivity limit of images are $\sim10$--$20\arcsec$ which correspond
to linear sizes of the galaxies $\sim 10$--$20$~kpc.

In Fig.~\ref{corr} we present correlation between X-ray luminosity and
luminosity of AGN in $O[III],\lambda 5007$. As it was shown by different
authors \cite[see e.g.,][]{heckman05} these two quantities are well
correlated.

Our measurement of AGN luminosities in X-rays and oxygen line are
consistant with these results, taking into account possible
variability of AGN luminosity on time scales of months and years
between our X-ray and optical measurements.

\begin{figure*}[t]
\hbox{
\includegraphics[width=0.8\columnwidth]{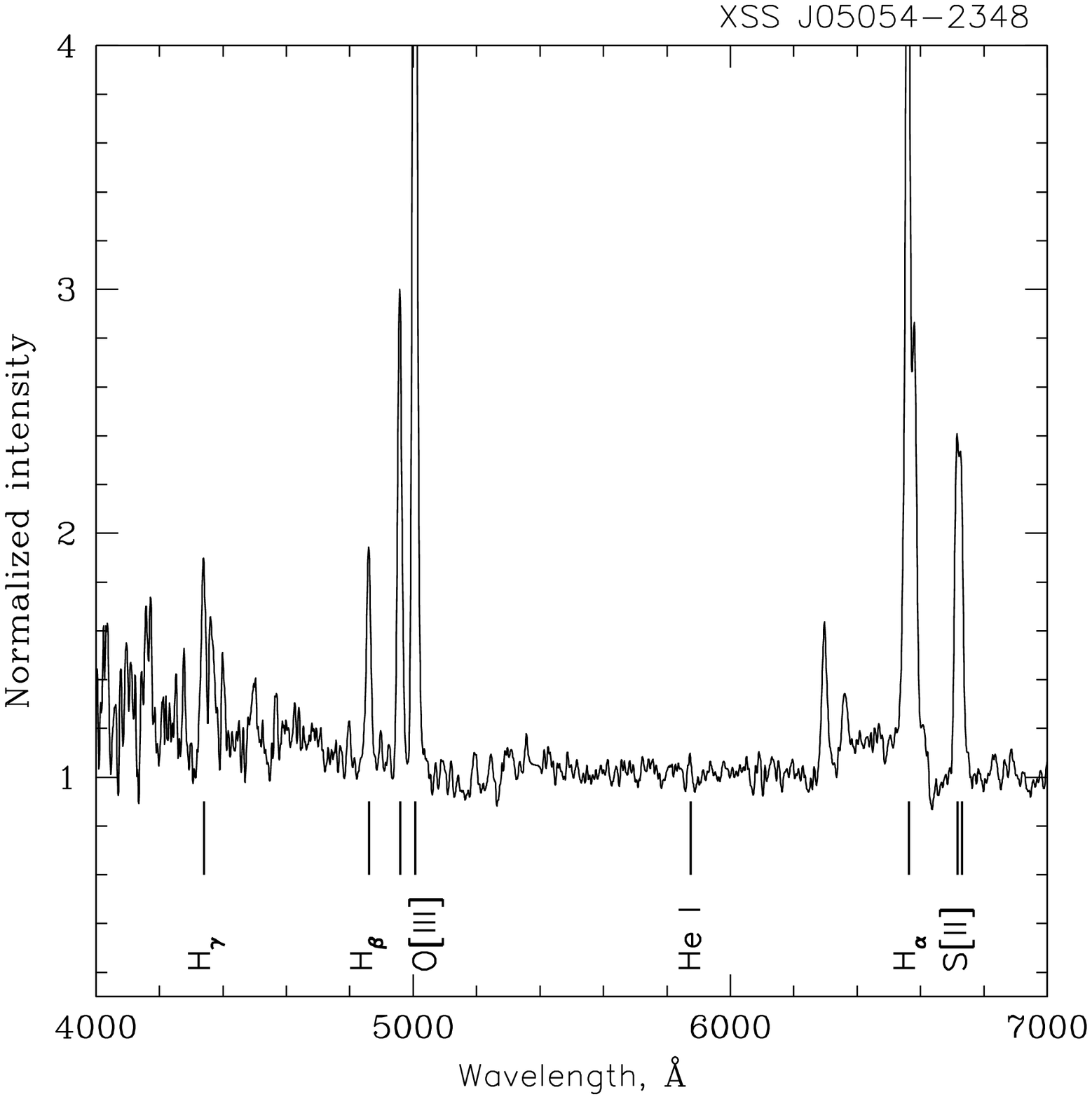}
\includegraphics[width=0.8\columnwidth]{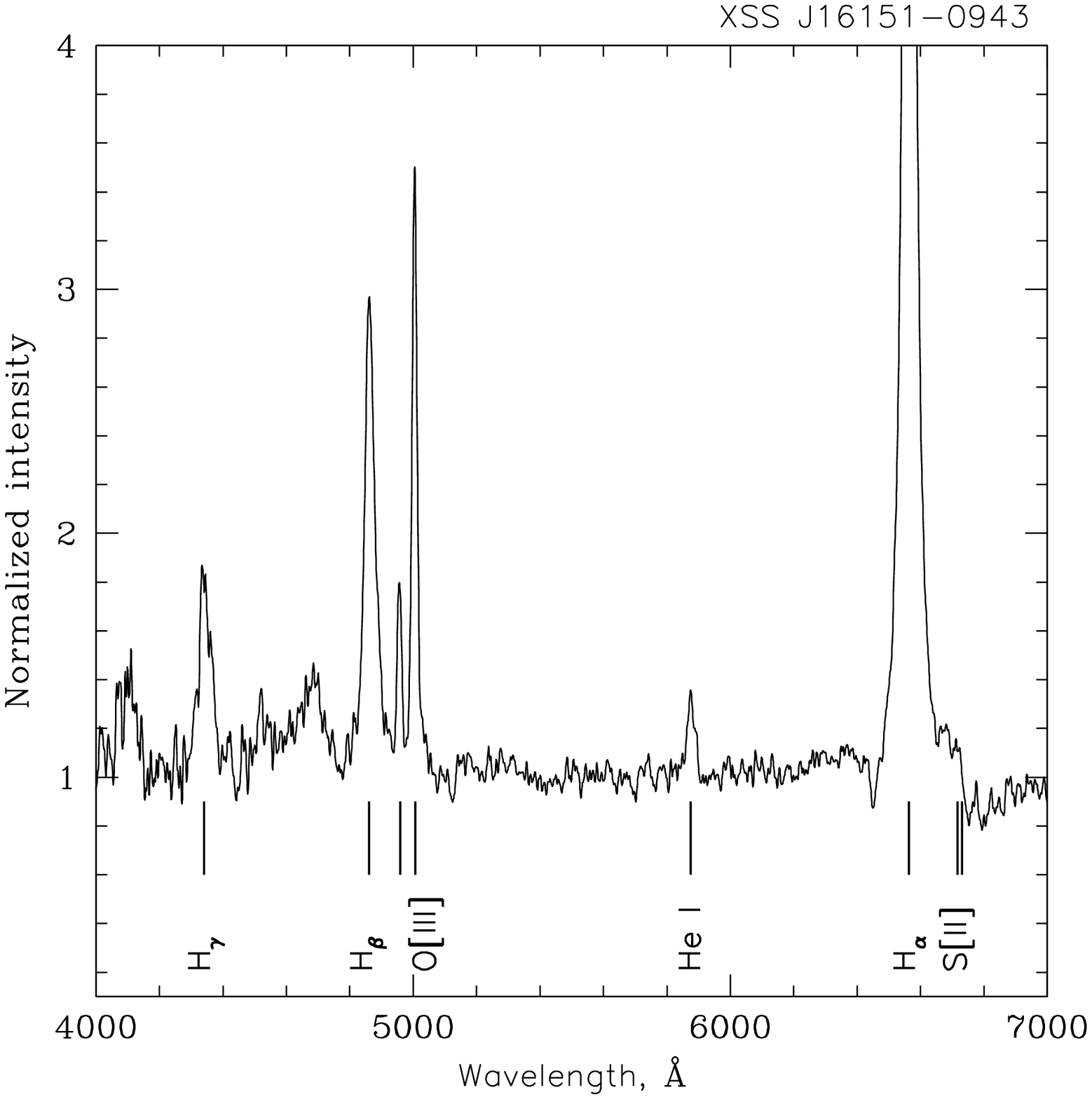}
}
\hbox{
\includegraphics[width=0.8\columnwidth]{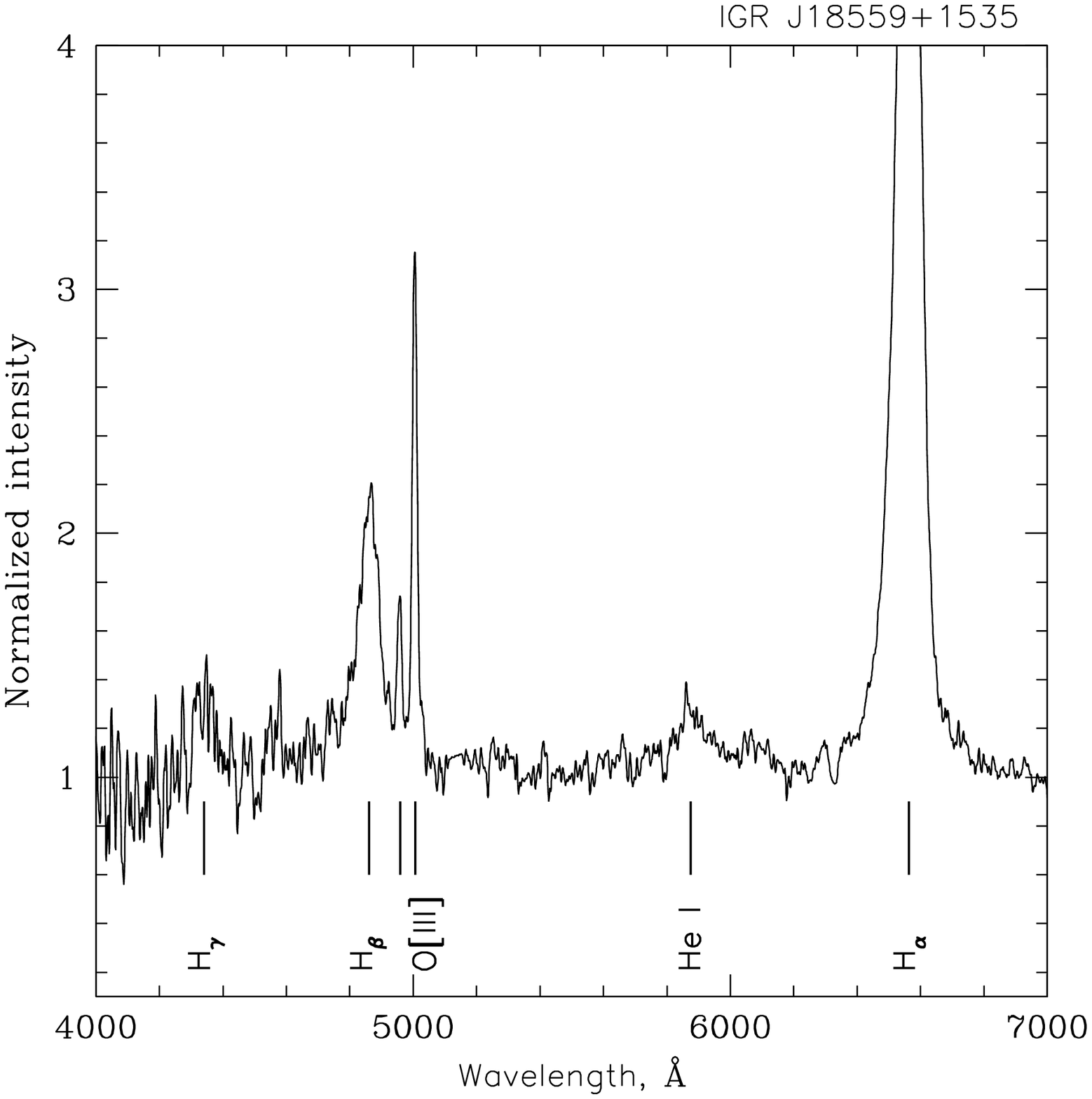}
\includegraphics[width=0.8\columnwidth]{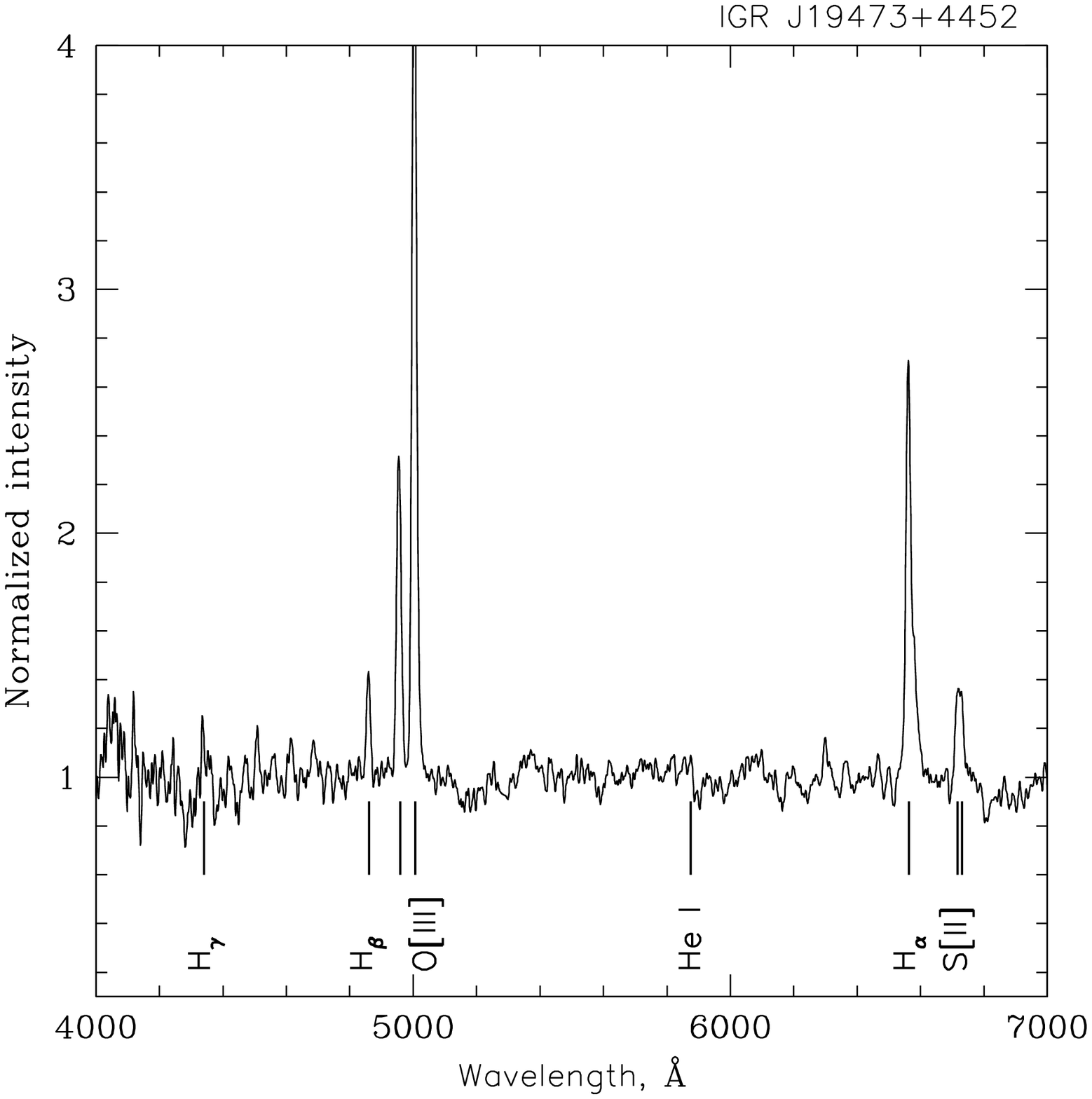}
} \hbox{
\includegraphics[width=0.8\columnwidth]{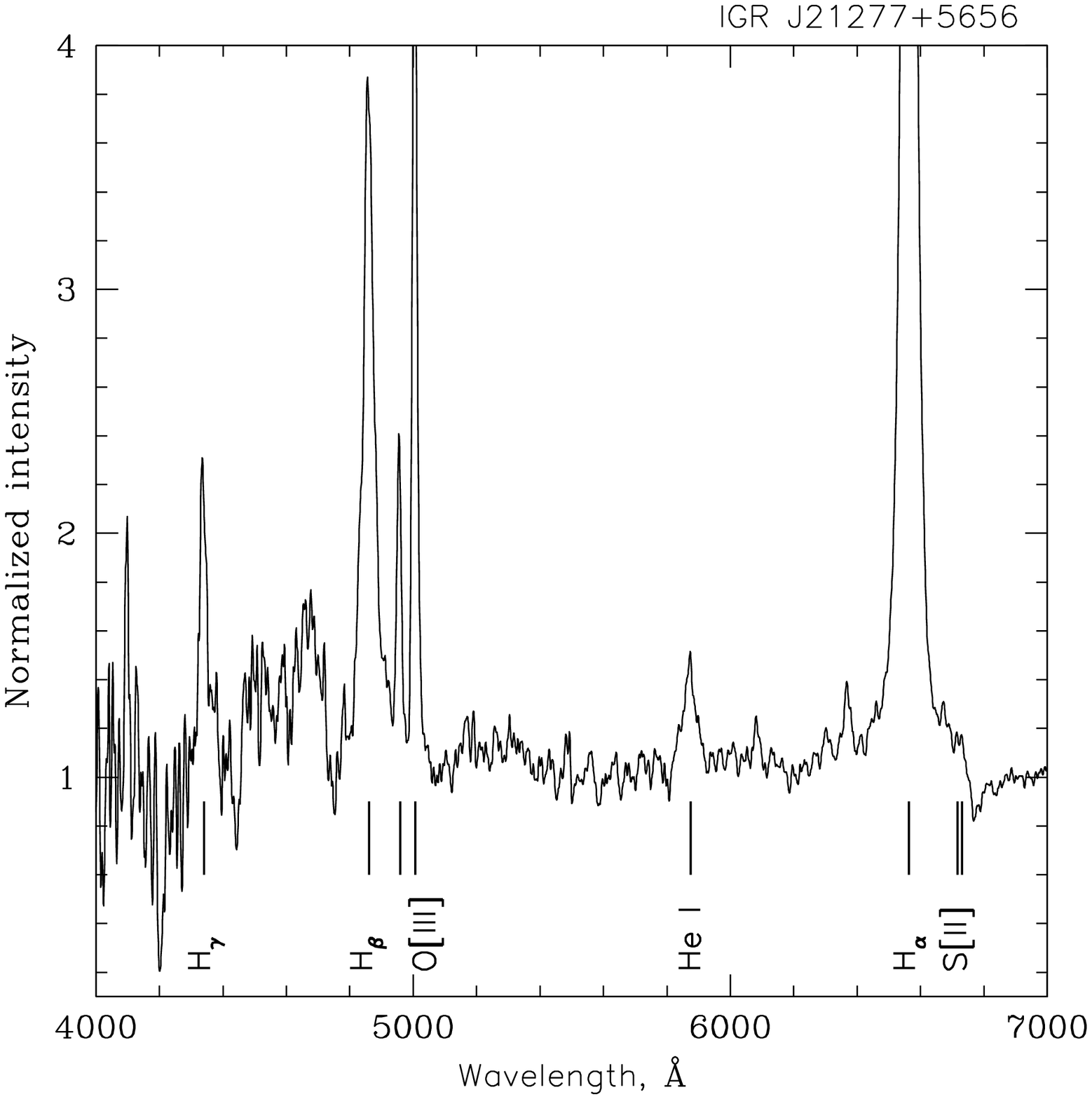}
\includegraphics[width=0.8\columnwidth]{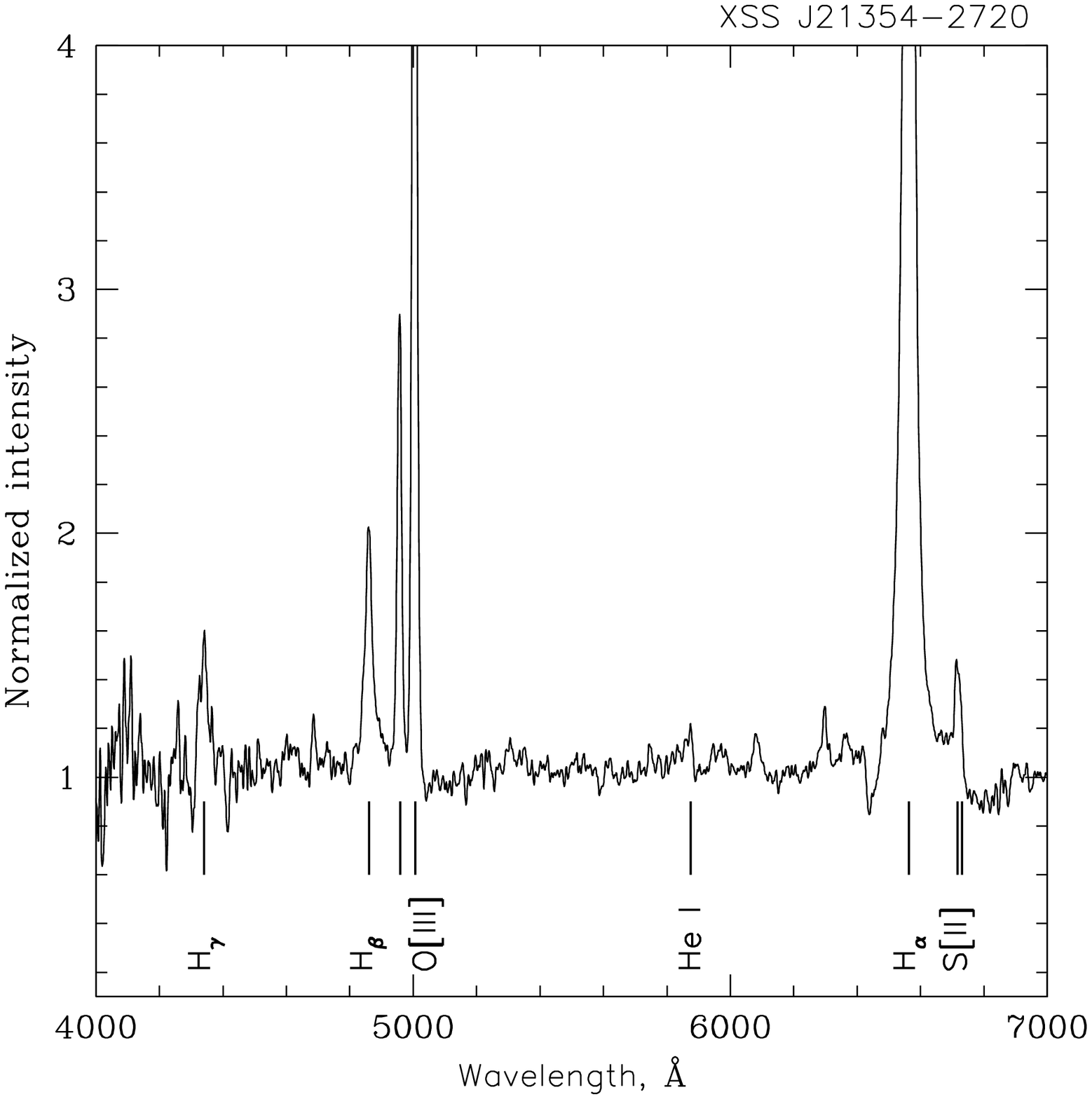}
} \caption{Normalized spectra of identified AGNs.\label{spectra}}
\label{spectra}
\end{figure*}

\begin{figure}[htb]
\includegraphics[width=\columnwidth]{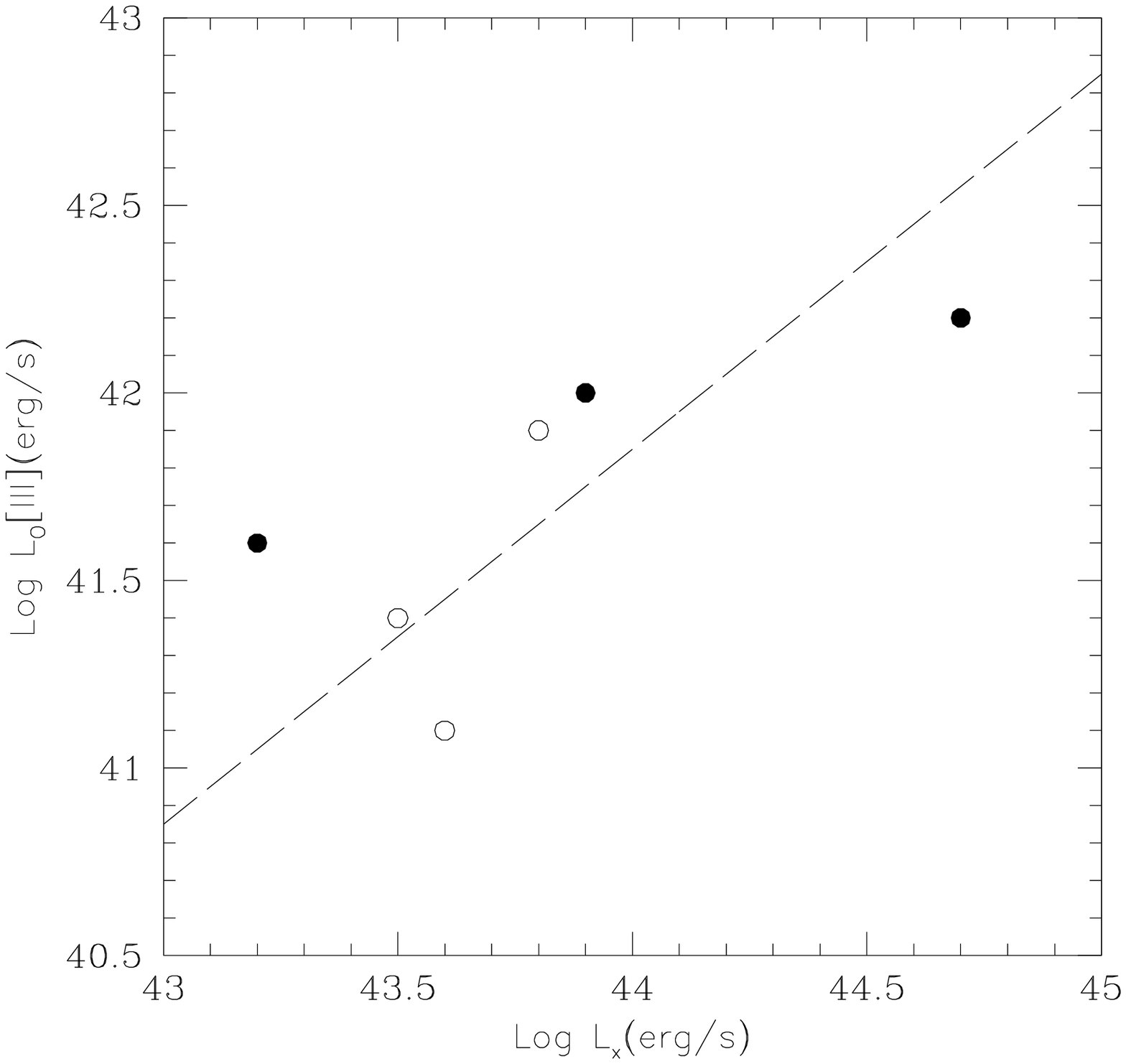}
\caption{Correlation of optical luminosity of AGNs in emission line
  O[III]$\lambda 5007$ with X-ray luminosity. Open circles show type 2
  Seyferts, filled circles --- type 1 Seyferts. Dashed line shows linear
  correlation with $\log L_{\rm x}/L_{\rm O[III]}\sim 2$\label{corr}}
\label{corr}
\end{figure}

\section*{CONCLUSION}

We have performed series of observations of AGN candidates discovered by
RXTE and INTEGRAL observatories with RTT150 telescope. Using low resolution
spectrometer TFOSC we obtained a number of spectra within localization
regions of these X-ray sources. In all cases we have discovered relatively
bright ($Rc\la17$) optical objects with redshifted emission lines, typical
for AGNs. We present redshifts of these AGNs and main parameters of their
emission lines, which allowed us to classify them as Seyfert 1 or Seyfert 2.
Six newly identified AGNs in local Universe is a significant contribution to
the existing samples of nearby AGNs. Therefore, further continuation of our
campaign of identification of RXTE and INTEGRAL AGN candidates in optical is
an important task which will help to improve our knowledge of demography of
supermassive black holes in local Universe.

{\it Authors are gratefull to the Director of T\"UBITAK National observatory
  (Turkey) Prof. Zeki Aslan for his strong support of this work. Authors
  thank A.Tkachenko (IKI RAN), A.Galeev and R.Zhuchkov (KSU) for their help
  in performing of photometric observations and also to Prof. N.Sakhibullin
  to his useful discussions of obtained results. Work was supported by
  grants RFFI 05-02-17744, NSH-1789-2003.2 and NSH-2083.2003.2 and by
  program of Presidium of the Russian Academy of Sciences ``Variable
  phenomena in astronomy''.}

{\it NOTE: When paper was already accepted we have noticed that one AGN of
  our sample (IGR J18559+1535) was also identified by the group of
  \cite{masetti05} }

\end{document}